\documentclass[aps,pra,reprint,groupedaddress]{revtex4-2}

\usepackage{amssymb,bbold}
\usepackage{amsmath}
\usepackage{amssymb}
\usepackage[utf8]{inputenc}
\usepackage[english]{babel}
\usepackage{amssymb}
\usepackage{amsthm}
\usepackage{bbm}
\usepackage{graphicx}
\usepackage{hyperref}
\usepackage{physics}
\usepackage{systeme}
\usepackage{times}
\usepackage{epstopdf}
\usepackage{float}
\usepackage{ulem}
\RequirePackage{color}
\RequirePackage[T1]{fontenc}
\RequirePackage{graphicx}
\RequirePackage{mathptmx}
\RequirePackage{flushend}

\newcommand{\orcid}[1]{\href{https://orcid.org/#1}
	{\includegraphics[width=7pt]{orcid.png}}}
\hypersetup{
	colorlinks=true,linkcolor=blue,citecolor=blue,
	filecolor=blue,urlcolor=blue,breaklinks=true
}

\begin{document}

\title{Optical and electronic properties of a two-dimensional quantum ring under rotating effects}

\author{Daniel F. Lima}
\email[Daniel F. Lima - ]{daniel.franca.lima@gmail.com}
\affiliation{Departamento de F\'{\i}sica, Universidade Federal do Maranh\~{a}o, 65085-580 S\~{a}o Lu\'{\i}s, MA, Brazil}
	
\author{Frankbelson dos S. Azevedo}
\email[Frankbelson dos S. Azevedo - ]{frfisico@gmail.com}
\affiliation{Departamento de F\'{\i}sica, Universidade Federal do Maranh\~{a}o, 65085-580 S\~{a}o Lu\'{\i}s, MA, Brazil}
	
\author{Lu\'{i}s Fernando C. Pereira}
\email[Lu\'{i}s Fernando C. Pereira - ]{luisfernandofisica@hotmail.com}
\affiliation{Departamento de F\'{\i}sica, Universidade Federal do Maranh\~{a}o, 65085-580 S\~{a}o Lu\'{\i}s, MA, Brazil}
	
\author{Cleverson Filgueiras}
\email[Cleverson Filgueiras - ]{cleverson.filgueiras@ufla.br }
\affiliation{Departamento de F\'{i}sica, Universidade Federal de Lavras, Caixa Postal 3037, 37200-000, Lavras, Minas Gerais, Brazil}
	
\author{Edilberto O. Silva}
\email[Edilberto O. Silva - ]{edilberto.silva@ufma.br}
\affiliation{Departamento de F\'{\i}sica, Universidade Federal do Maranh\~{a}o, 65085-580 S\~{a}o Lu\'{\i}s, MA, Brazil}

\date{\today}
	
\begin{abstract}
This work presents a study on the nonrelativistic quantum motion of a charged particle in a rotating frame, considering the Aharonov-Bohm effect and a uniform magnetic field. We derive the equation of motion and the corresponding radial equation to describe the system. The Schrödinger equation with minimal coupling incorporates rotation effects by substituting the momentum operator with an effective four-potential. Additionally, a radial potential term, dependent on the average radius of the ring, is introduced. The analysis is restricted to motion in a two-dimensional plane, neglecting the degree of freedom in the $z$-direction. By solving the radial equation, we determine the eigenvalues and eigenfunctions, allowing for an explicit expression of the energy. The probability distribution is analyzed for varying rotating parameter values, revealing a shift of the distribution as the rotation changes, resulting in a centrifugal effect and occupation of the ring's edges. Furthermore, numerical analysis demonstrates the significant rotational effects on energy levels and optical properties, including optical absorption and refractive coefficients.
\end{abstract}
	
\maketitle

\section{Introduction}

The behavior of Two-Dimensional Electron Gas (2DEG) in quantum rings at the mesoscopic scale, under the influence of specific types of confinement potential, Aharonov-Bohm potential, and external magnetic fields, has been extensively investigated. These conditions give rise to various physical phenomena that affect the electronic properties of the system. Notably, the Landau levels are modified due to the influence of a radial potential, leading to an increase in magnetization, and persistent current \cite{PRB.1999.60.5626,PRB.2004.69.195313}. Additionally, quantum rings exhibit non-linear optical properties, as demonstrated through calculations of the Optical Absorption Coefficients (OACs) and Refractive Index Changes (RICs). These properties can be influenced by external magnetic fields, Confinement Potential, and Aharonov-Bohm effect \cite{liang2011optical,duque2012quantum,xie2013optical}. Quantum rings remain an active and open field of exploration, attracting significant interest and extensive research due to their physical background and applications. These include the study of effects on optical properties of cylindrical quantum dots influenced by temperature and pressure \cite{jaouane2022effects}, calculation of optical properties of spherical quantum dots with combined confining potentials \cite{duan2022calculation}, and search of optical properties of quantum disk with azimuthal distortion that may be beneficial in designing of optoelectronic devices \cite{talwar2022optical}. Applications of quantum rings also include the utilization of quantum rings in nano-flash memories \cite{nowozin2013self}, photonic devices \cite{michler2003single}, qubits for spintronic quantum computing \cite{PhysRevB.74.125426,szopa2010flux}, as well as magnetic random access memory, recording medium, and other spintronic devices \cite{fomin2013physics}.

Investigating rotational effects in quantum mechanics has a long history of extensive research. For instance, the Coriolis force in a rotating frame of reference is known to be analogous to the Lorentz force on a charged particle due to a magnetic field \cite{leighton1965feynman}. Tsai and Neilson \cite{tsai1988new} also showed that the quantum interference effect in rotating frames is analogous to the well-known Aharonov-Bohm effect \cite{PR.1959.115.485}. Gaining a deeper comprehension of how rotation influences quantum systems offers valuable insights into the fundamental behavior of particles. However, investigating rotational effects in systems like quantum rings is a recent topic. Analytical investigations have recently addressed the rotational effects in quantum ring systems \cite{dantas2015quantum,fonseca2016rotating}.
Furthermore, recent numerical investigations have showcased notable rotational effects on electronic states, persistent currents, and magnetization in quantum rings \cite{pereira20221d,pereiramodification}. Overall, the impact of rotating in the study of quantum rings has interesting implications on energy levels and the fundamental quantum behavior of the system. Therefore, studying the spatial distribution of particles and optical properties of quantum rings under rotational effects is still an open topic. Future investigations in this topic advance our theoretical understanding and hold promise for progressing areas such as quantum optics, optoelectronics, and their applications.

From this standpoint, this work investigates the impact of rotation on the nonrelativistic quantum motion of a charged particle within a rotating frame, explicitly considering the Aharonov-Bohm effect and a uniform magnetic field. In this study, we derive the equation of motion and the corresponding radial equation to describe the system. The analysis employs the Schrödinger equation with minimal coupling, where rotation effects are incorporated by substituting the momentum operator with an effective four-potential. We introduce a radial potential term that depends on the average radius of a ring. While not included in this manuscript, the potential under consideration also enables the study of optical properties, such as conductance measurement \cite{PRB.1996.53.6947} and photoionization cross-section \cite{SM.2013.58.94}. It is important to note that the present investigation is limited to two-dimensional motion, neglecting the degree of freedom in the z-direction. By determining the eigenvalues and eigenfunctions, our objective is to unravel the intricate relationship between rotation and quantum behavior. The probability distribution is analyzed for various values of the rotating parameters. Furthermore, we perform numerical analysis to investigate the rotational effects on energy levels and optical properties, including optical absorption and refractive coefficients. These are crucial in understanding the system's behavior under rotation. 

This article is organized as follows: Section \ref{EM} provides an in-depth analysis of the equation of motion, presenting its solution and introducing the concept of the quantum ring. The derivation of the equation of motion and its implications for the system under rotation is discussed in detail. In Section \ref{results}, we present the numerical results obtained from the implementation of rotations and provide a comprehensive discussion of these results. The effects of rotation on energy levels, optical properties, and probability distributions are examined, shedding light on the intricate relationship between rotation and quantum behavior. Finally, in Section \ref{sec:conclusions}, we draw conclusions based on our findings and highlight the significance of the explored rotation effects in quantum mechanics.

\section{The Equation of Motion}
\label{EM}

\begin{figure}[!t]
\centering
\includegraphics[width=\columnwidth]{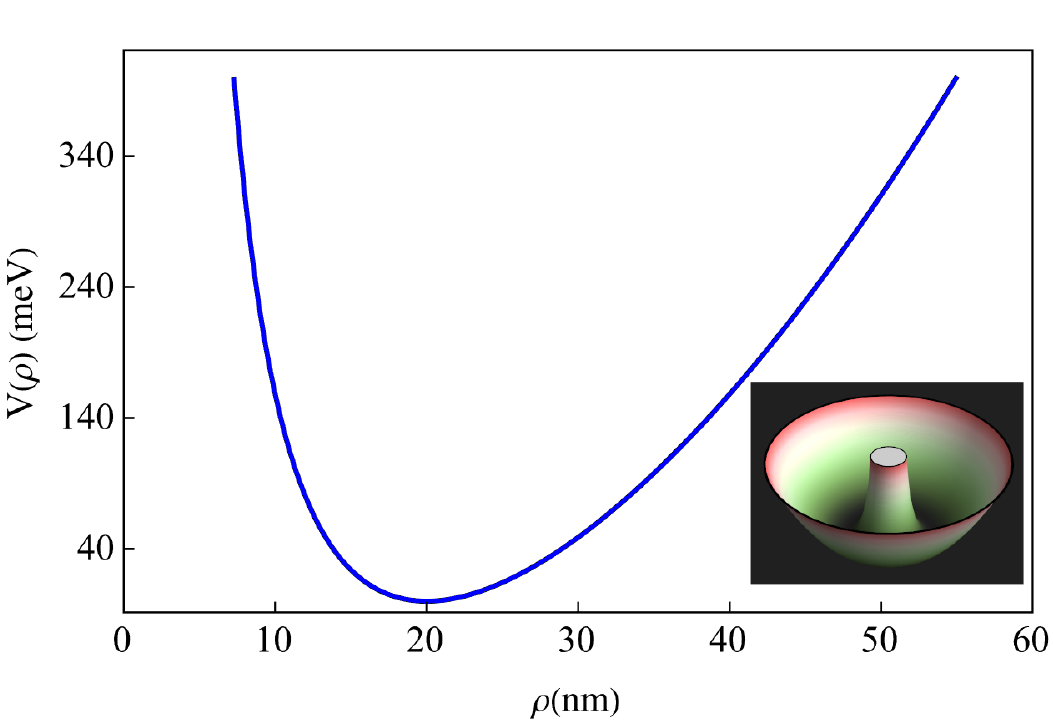}
\caption{Profile of the radial potential describing a ring of average radius $\rho_0=20 \text{~} \text{nm}$ and $\hbar\omega_0=40 \text{~} \text{meV}$.}
\label{f1}	
\end{figure}

In this section, we study the nonrelativistic quantum motion of a charged
a particle of mass $m_{e}$ in a rotating frame in the presence of both the Aharonov-Bohm effect and uniform magnetic field without considering the particle's spin effects. We shall obtain the equation of motion and its corresponding radial equation of motion. Subsequently, we must solve it and determine the particle's eigenenergies and wave functions. Evidently, for our purposes, we shall only consider the Schrödinger equation with minimal coupling, which we follow the model presented in Chapter 17 of Ref. \cite
{Book_Rotating_Frames}. The rotational and electromagnetic effects are included in the Schrödinger equation by substituting 
\begin{equation}
p^{\upsilon}\rightarrow p^{\upsilon}-\mu \,A_{\text{eff}}^{\upsilon}, \;\;(\upsilon =0,1,2,3),
\end{equation}
where $A_{\text{eff}}^{\upsilon}$ is effective four-potential, which
includes the Electromagnetic four-potential $A_{\text{ele}}^{\upsilon}$, and the gauge
field for the rotating frame $A_{\text{rot}}^{\upsilon}=\left( A_{\text{rot}}^{0},\mathbf{A}
_{\text{rot}}\right) $. The electromagnetic four-potential is specified by $
A_{\text{ele}}^{\upsilon}=\left( A^{0}_{\text{ele}},\mathbf{A}_{\text{ele}}=\mathbf{A}_{1}+\mathbf{A}_{2}\right)$. As we shall specify below, the effective vector potential $
\mathbf{A}_{\text{ele}}$ is given in terms of the Aharonov-Bohm potential and the
potential due to a uniform magnetic field in the $z$-direction, with $\mathbf{\nabla}\cdot \mathbf{A}_{\text{ele}}=0$ and $A^{0}_{\text{ele}}=0$. We define the magnetic flux tube for the Aharonov-Bohm problem as 
\begin{equation}
e\mathbf{A}_{1}=\left(0,-\frac{\phi}{\rho},0\right),\;\;e\mathbf{B}_{1}=\left(0,0,-\phi \frac{\delta (\rho)}{\rho}\right), \label{gf}
\end{equation}
The quantity $\phi =e\Phi /2\pi \hbar =\Phi /\Phi_{0}$ is related to the
Aharonov-Bohm flux, in which $\Phi $ denotes the magnetic flux and $\Phi
_{0}=h/e$ indicates the quantum of magnetic flux. The potential vector that
generates the uniform magnetic field in the $z$-direction is specified by
\begin{equation}
\mathbf{A}_{2}=\left( 0,\frac{1}{2}B\rho ,0\right) \mathbf{,~~\ B}
_{2}=\left( 0,0,B\right) .
\end{equation}
The third configuration characterizes the gauge field for the rotating frame
\begin{equation}
A_{\text{rot}}^{\upsilon}=\left( -\frac{1}{2}\left( \boldsymbol{\Omega }\times \mathbf{r
}\right) ^{2},\boldsymbol{\Omega }\times \mathbf{r}\right) ,
\end{equation}
where $\boldsymbol{\Omega}$ is the angular velocity. Here, we assume that $\boldsymbol{\Omega}$ has only the $z$ component so that in cylindrical
coordinates, the coordinate vector reads $\mathbf{r}=\rho \,\boldsymbol{\hat{\rho}}$, and consequently, we write $\boldsymbol{\Omega }\times \mathbf{r}=\Omega \rho \,\boldsymbol{\hat{\varphi}}$. Furthermore, it is well known in the literature that the Aharonov-Bohm problem has translational invariance in the $z$-direction, so we can exclude the $z$ degree of freedom by imposing $p_{z}=z=0$ \cite{PRL.1990.64.503,PRD.2012.85.041701,AoP.2013.339.510}.
Thus, we shall study the motion of the electron constrained to move only in the plane. Therefore, we must study the combined effects involving the rotation $\boldsymbol{\Omega}$, the potentials $\mathbf{A}_{1}$ and $\mathbf{A}_{2}$, and the field $\mathbf{B}_{2}$. The equation of motion to be solved is written as
\begin{equation}
\frac{1}{2 \mu }\left( \mathbf{p}-e\mathbf{A}_{\text{ele}}-\mu\boldsymbol{\Omega}\times \mathbf{r}\right)^{2}\psi -\frac{1}{2}\mu(\boldsymbol{\Omega}
\times \mathbf{r})^{2}\psi +V\left( \mathbf{r}\right)\psi=E\psi,   \label{se}
\end{equation}
where $V\left(\mathbf{r}\right)$ represents the radial potential of a 2D ring defined by (see Figure \ref{f1})
\begin{equation}
V\left(\rho\right) =\frac{\mu \omega _{0}^{2}}{8}\left({\rho}-\frac{\rho_{0}^{2}}{\rho}\right)^2,\label{vr}
\end{equation}
where $\rho_{0}$ defines the average radius of the ring and also represents the point where the minimum potential is located. The quantity $\omega_{0}$ characterizes the strength of the transverse confinement. Besides characterizing an exactly soluble model, the radial potential (\ref{vr}) describes a localized ring of finite width, providing a convenient theoretical tool to study electronic states, as well as their dependence on the magnetic field in a 2D
ring \cite{PRB.1999.60.5626,PRB.2004.69.195313}. 

After replacing the field and potential configurations in Eq. (\ref{se}), we
obtain
\begin{align}
& \frac{1}{\rho }\frac{\partial }{\partial \rho }\left( \rho \frac{\partial
\psi }{\partial \rho }\right) -\frac{1}{\rho ^{2}}\left[ \left( \frac{
\partial }{i\partial \varphi }-\phi \right) ^{2}+  \frac{\mu^2 \omega_{0}^2 \rho_{0}^4}{4\hbar^2}\right] \psi \notag \\
&-\rho ^{2}\left( \frac{\mu ^{2}\omega _{c}^{2}}{4\hbar ^{2}}+
\frac{\mu ^{2}\Omega \omega _{c}}{\hbar ^{2}}+\frac{
\mu^2 \omega_{0}^2 \,}{4\hbar ^{2}}\right) \psi    \notag \\
& +\frac{\mu \omega _{c}}{\hbar }\frac{\partial \psi }{i\partial \varphi }
+\frac{2\mu \Omega
  }{\hslash } \frac{\partial \psi }{ i\partial \varphi } -\frac{2\mu \Omega
 \phi }{\hslash }\psi -\frac{\mu \omega _{c} \phi }{\hslash } \psi \notag \\
&+ \frac{\mu^2 \omega_{0}^{2}\rho _{0}^{2}}{2 \hbar^2}
\psi =-\frac{2\mu E}{\hbar ^{2}}\psi.
\end{align}
Assuming the eigenfunctions to be of the form 
\begin{equation}
\psi \left( \rho ,\varphi \right) =f(\rho )e^{im\varphi },  \label{wavef}
\end{equation}
where $m=0,\pm 1,\pm 2,\pm 3,\ldots $ is the angular momentum quantum
number, we obtain the radial equation 
\begin{equation}
\frac{1}{\rho }\frac{d}{d\rho }\left( \rho \frac{dF}{d\rho }\right) -\frac{
L_{m}^{2}}{\rho ^{2}}F-\frac{\mu ^{2}\varpi ^{2}\rho ^{2}}{4\hbar ^{2}}F+k^{2}F=0,
\label{re}
\end{equation}
where $\lambda =\sqrt{\hbar /\mu \varpi }$ is the effective magnetic length
renormalized by the rotation, $L_{m}=\sqrt{\left( m-\phi \right)^{2}+ \mu^2 \omega_{0}^2 \rho_{0}^4/4\hbar^2}$ is the effective angular momentum, $\varpi =\sqrt{\omega
_{c}^{2}+4\Omega \omega _{c}+\omega_{0}^2 }$ is the effective cyclotron
frequency,~$\omega _{c}=eB/\mu $ is the cyclotron frequency, and $k^{2}=\mu
\left( 2\Omega +\omega _{c}\right) \left( m-\phi \right) /\hbar +2\mu \left(
\mu \omega _{0}^{2}\rho _{0}^{2}/4+{E}_{nm}\right) /\hbar ^{2}$. Solving (\ref{re}), we obtain
the following eigenvalues and eigenfunctions:
\begin{align}
\psi_{nm}\left( \rho ,\varphi \right) &=\frac{1}{\lambda }\sqrt{\frac{\Gamma
\left( n+L_{m}+1\right) }{2^{L_{m}+1} n!\left[ \Gamma \left( L_{m}+1\right) 
\right] ^{2}\pi }}\left( \frac{\rho }{\lambda }\right) ^{L_{m}}e^{im\varphi
}e^{-\frac{\rho ^{2}}{4\lambda ^{2}}}\notag \\ & \times~_{1}F_{1}\left( -n,L_{m} +1,\frac{\rho ^{2}}{2\lambda ^{2}}\right) ,  \label{ef}
\end{align}
\begin{equation}
{E}_{nm}=\frac{\hbar ^{2}}{2\mu \lambda ^{2}}\left(
2n+L_{m}+1\right) -\frac{\hbar }{2}\left( 2\Omega +\omega _{c}\right) \left(
m+\phi \right)-\frac{\mu }{4}
\omega _{0}^{2}\rho _{0}^{2}.
\label{eny}
\end{equation}
It is important to express energy more explicitly as 
\begin{align}
&{E}_{nm}=\left( n+\frac{1}{2}\sqrt{\left( m-\phi \right) ^{2}+\frac{
\mu ^{2}\omega _{0}^{2}\rho _{0}^{4}}{4\hbar ^{2}}}+\frac{1}{2}\right) \hbar \notag \\
&\times\sqrt{\omega _{c}^{2}+\omega _{0}^{2}+4\Omega \omega_{c}}-\frac{\hbar }{2}
\left( 2\Omega +\omega_{c}\right) \left( m-\phi \right) -\frac{\mu }{4}
\omega _{0}^{2}\rho_{0}^{2}.\label{enye}
\end{align}
In the absence of rotating effects, we obtain
\begin{align}
&{E}_{nm}=\left( n+\frac{1}{2}\sqrt{\left( m-\phi \right) ^{2}+\frac{
\mu ^{2}\omega_{0}^{2}\rho_{0}^{4}}{4\hbar ^{2}}}+\frac{1}{2}\right) \hbar \notag\\
&\times \sqrt{\omega_{c}^{2}+\omega_{0}^{2}}-\frac{\hbar \omega _{c}}{2}\left(
m-\phi \right)-\frac{\mu }{4}\omega_{0}^{2}\rho _{0}^{2},
\end{align}
which recovers the result in Reference \cite{PRB.1999.60.5626}, as expected.

\section{Results and discussion}
\label{results}

In this section, we numerically investigate the results obtained in Section \ref{EM} for the case of a 2D quantum ring with an average radius of $\rho_0 = 20$ nm and $\hbar\omega_0 = 40$ meV. We use the data collected from a 2D GaAs heterostructure, which can easily be found in the literature. We complete our analysis by studying the impact of rotation on the linear and nonlinear optical properties of an electron in a 2D quantum ring. All calculations were performed using the following physical parameters \cite{PRB.1999.60.5626,duque2012quantum}:
 $\mu = 0.067 \,\mu_e$, where  $\mu_e=9.1094\times10^{-31}\,eV/c^{2}$ is the electron mass, $n_{r}=3.2$ is the refractive index, $\epsilon_0=8.854\times10^{-12}\,\text{F/m}$ is the permittivity of free space and  $c=2.99\times10^{8}\,m/s$ is the speed of light. We have also considered the specific values: the incident optical intensity $I=40$ GW/m$^{2}$, the relaxation time $\Gamma_{0}=1/0.2\, \text{ps}$, the density of states $\sigma_\nu=5.0\times10^{22}\,m^{-3}$ and the permeability of the system $\mu_0=4\pi\times10^{-7}\,\text{T m/A}$ \cite{shao2011third,liang2011optical}.

\begin{figure}[!h]
\centering
\includegraphics[width=\columnwidth]{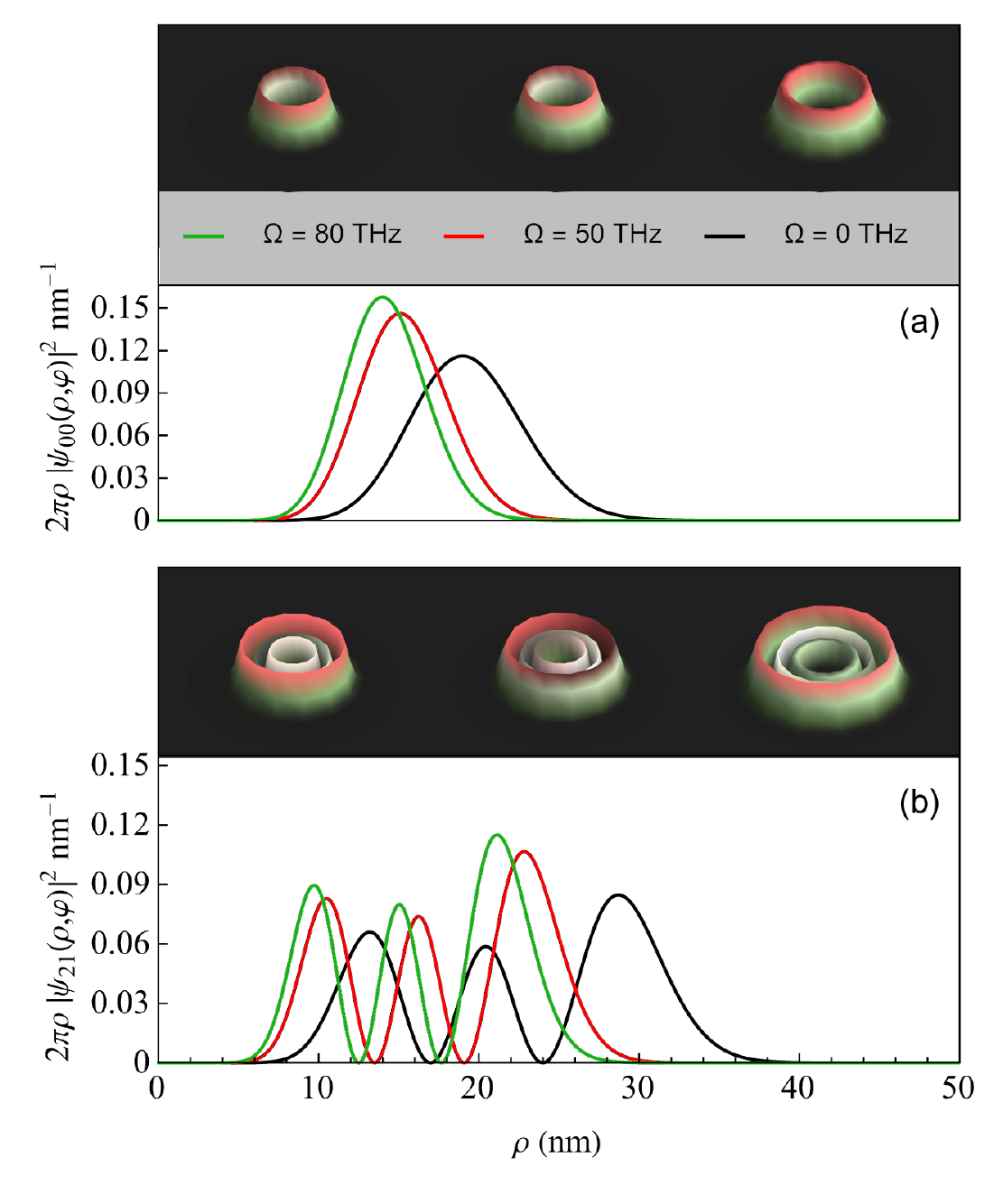}
\caption{Probability distribution function for positive rotations. (a) For ($n=0,m=0$). (b) For ($n=2,m=1$).}
\label{FigAbs_Refr}	
\end{figure}

Using the normalized eigenfunction (\ref{ef}), we can evaluate the probability distribution of finding the particle in certain states $n$ and $m$. This allows us to analyze the system's behavior and understand its properties. Figure \ref{FigAbs_Refr} illustrates how the probability distribution changes with a positive rotating parameter for the states with ($n=0,m=0$) and ($n=2,m=1$) (on the top, we have a 3D visualization of the probability profile). As $\Omega$ increases, the distribution shifts towards lower values, indicating that the particle is more likely to be found closer to the inner radius of the ring (inner edge) for higher rotations. The probability amplitude becomes more concentrated with less spacing between adjacent values as $\Omega$ increases. Additionally, the eigenfunction occupies the edges of the ring as $m$ increases giving rise to a centrifugal effect, with the particle having greater probability to be find at the outer edge.
For $n=2$, the amplitude of the third wave peak exceeds that of the preceding ones as we move along the horizontal axis. We then observe a symmetrical increase in the amplitudes that were previously decreasing. This phenomenon appears to occur for all $n$ greater than or equal to $1$. This behavior is expected since the potential in equation (\ref{vr}) approximates that of a pure Quantum Harmonic Oscillator \cite{Book.2005.Griffiths} for larger values of $\rho$ (i.e., greater outer radius), see Figure \ref{f1}.

\begin{figure}[!h]
\centering
\includegraphics[width=\columnwidth]{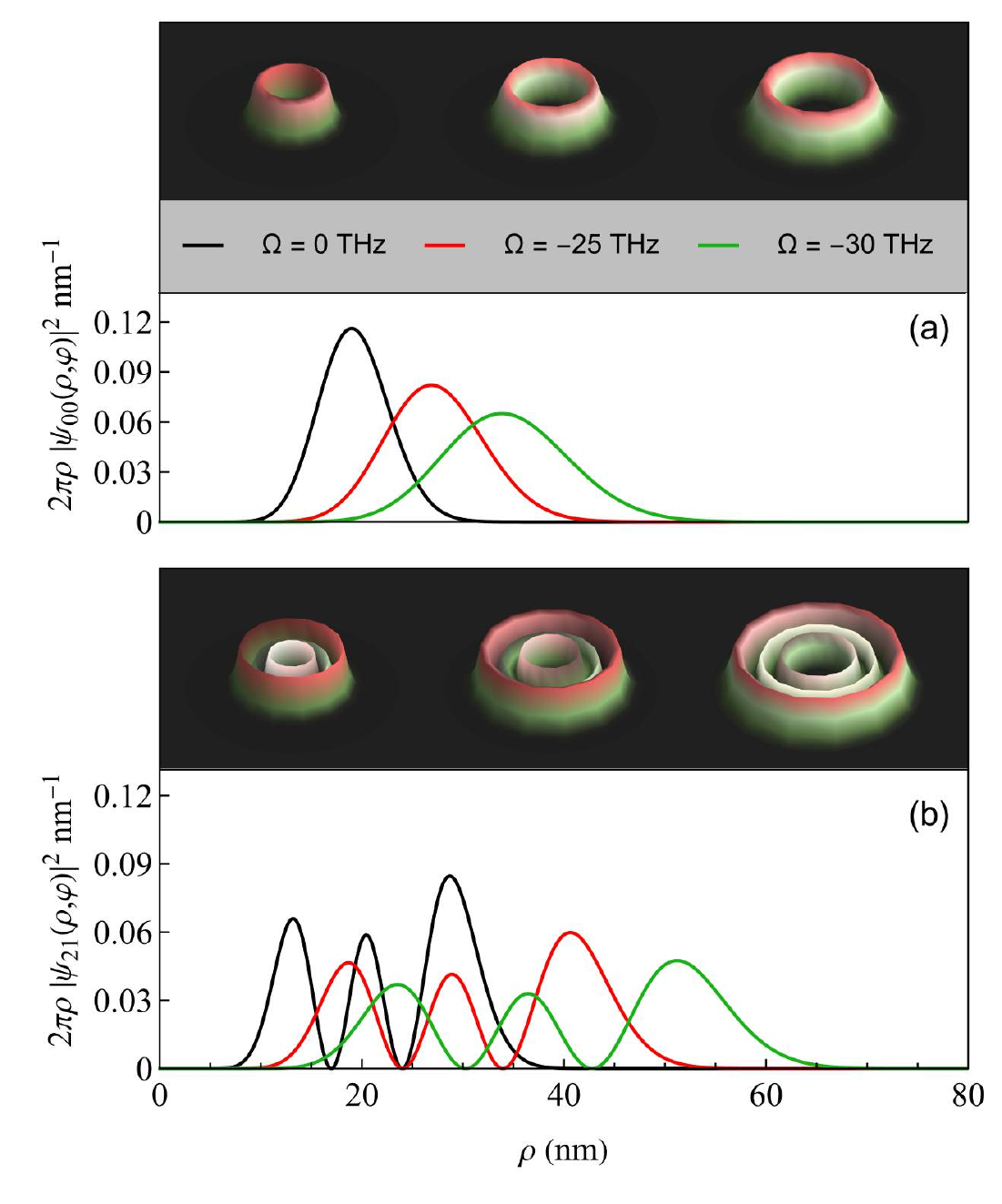}
\caption{The same as in Fig. \ref{FigAbs_Refr}, but considering negative rotations. In both Figures, we consider $\phi=0.5\,(h/e)$ and $B=15$ Tesla.}
\label{FigAbs_Refr2}	
\end{figure}

Similarly to the behavior observed in positive rotations, Figure \ref{FigAbs_Refr2} demonstrates how the probability distribution changes when negative rotating parameters are applied to states with ($n=0, m=0$) and ($n=2, m=1$) (on the top we have a 3D visualization of the probability profile). As the rotation rate $\Omega$ decreases, the distribution shifts towards higher values, indicating that the particle is more likely to be found closer to the outer radius of the ring (outer edge) for lower rotation rates. The probability amplitude becomes less concentrated, with greater spacing between adjacent values as $\Omega$ decreases. As the value of $m$ increases, the eigenfunctions tend to occupy the edges of the ring, with greater probability to be find at the outer edge.

\begin{figure}[!h]
\centering
\includegraphics[width=\columnwidth]{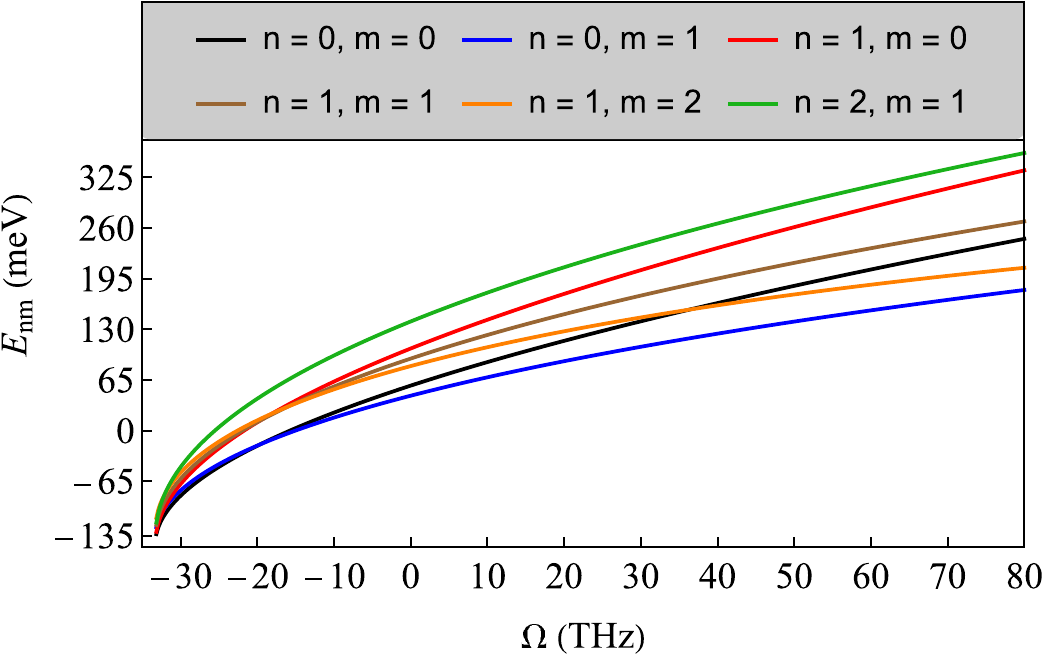}
\caption{Energy spectrum as a function of the rotation parameter for varying values of $n$ e $m$. We consider $\phi=0.5\,(h/e)$ and $B=15$ Teslas.}
\label{Energy_Rotation}	
\end{figure}

By examining the energy spectrum given by equation (\ref{enye}), we can see that the electronic states of the system are dependent on the rotation parameter (Figure \ref{Energy_Rotation}). For the observed states and range of rotation values considered, we find that the energy levels increase with the strength of the rotation. In the rotation range used in this work, we can see that the energy of the state $(n=0, m=0)$ can surpass states with larger quantum numbers as the value of $\Omega$ increases. For instance, in the range with positive rotations, the lowest energy level is not $(n=0, m=0)$. In this case, states with higher wave numbers like $(n=2, m=1)$ has a greater probability closer to the center of the ring due to the centrifugal effect caused by the quantum number $m$. Also, for the quantum ring we are studying here, the probability distribution of finding the electrons shows that the electrons are more likely to be found at the edges of the ring instead of being concentrated at the center of it (see Figures \ref{FigAbs_Refr} and \ref{FigAbs_Refr2}). This is expected as we consider various physical parameters in our calculations. However, the rotating parameter has a significant effect on the energy spectrum, as we can see from the states shown in Figure \ref{Energy_Rotation}. This rotational dependence on the energy spectrum is an important system feature and will be further analyzed in the context of Optical Properties.

By following the References \cite{duque2012quantum,ahn1987calculation}, in which the theory of the OACs and RICs are developed, we present the mathematics of optics necessary for our purposes in this work. In a two-level system approach with the dipole-allowed optical transitions being the states $\Delta m=\pm1$ \cite{yuan2018hydrogenic}, we choose the energy levels and the wavefunctions participating in the transition as:
\begin{equation}
 \psi_{1}=\psi_{00},\;\psi_{2}=\psi_{21}, \,\, \Delta E= E_{21}-E_{00}.
\end{equation}
Assuming the polarization of the incident radiation along the in-plane $x$-axis, the electric dipole transition matrix elements are 
\begin{equation}
M_{21}=\mel{\psi_2}{\rho \cos{\varphi}}{\psi_1},
\end{equation}
and considering that the electric dipole selection rules for $M_{11}$ and $M_{22}$ it will appear in terms of \(\int_{0}^{2 \pi} \cos\varphi \,d\varphi=0\) (due to the azimuthal symmetry of the system), hence
\begin{equation}
M_{11}=M_{22}=0.
\end{equation}
Considering it, the total optical absorption coefficient is given by 
\begin{equation} 
\alpha \left( \omega ,I\right) =\alpha ^{(1)}\left( \omega \right) +\alpha
^{(3)}\left( \omega ,I\right) ,  \label{at}
\end{equation}
with
\begin{equation}
\alpha ^{(1)}\left( \omega \right) =\hslash \omega \sqrt{\frac{\mu_0 }{
\epsilon _{r}}}\frac{\sigma _{\nu }\Gamma
_{21}\left\vert M_{21}\right\vert ^{2}}{\left(\Delta E-\hslash \omega \right) ^{2}+\left( \hslash \Gamma
_{21}\right) ^{2}}  \label{a1}
\end{equation}
\begin{figure}[!t]
\centering
\includegraphics[width=\columnwidth]{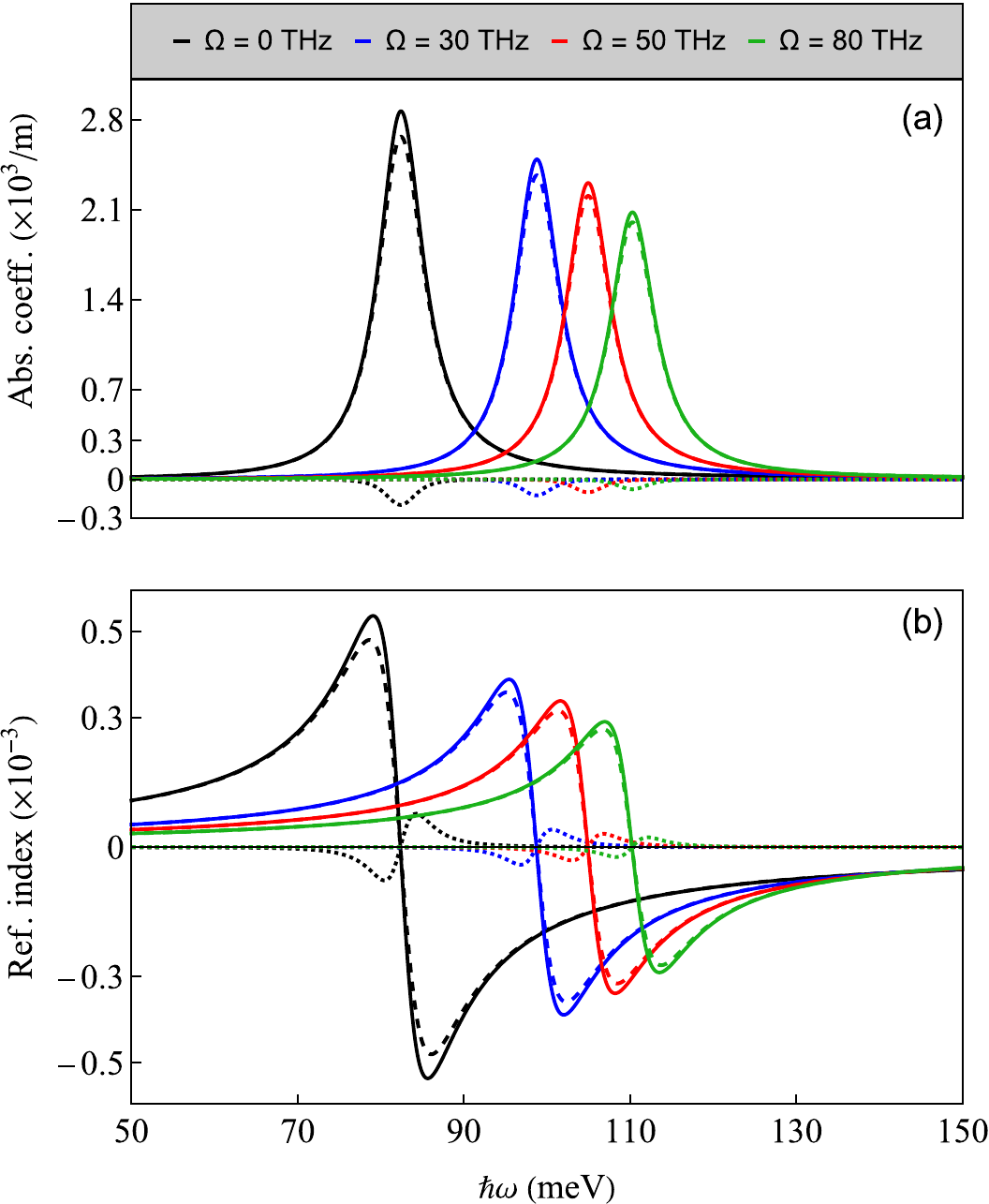}
\caption{(a) The linear (solid line), third-order non-linear (dotted line), and total absorption coefficient (dashed line) as a function  of the photon energy. (b) The linear (solid line), third-order non-linear (dotted line), and total refractive index (dashed line)  as a function of the photon energy for positive rotations.}
\label{f2}	
\end{figure}
\begin{figure}[!h]
\centering
\includegraphics[width=\columnwidth]{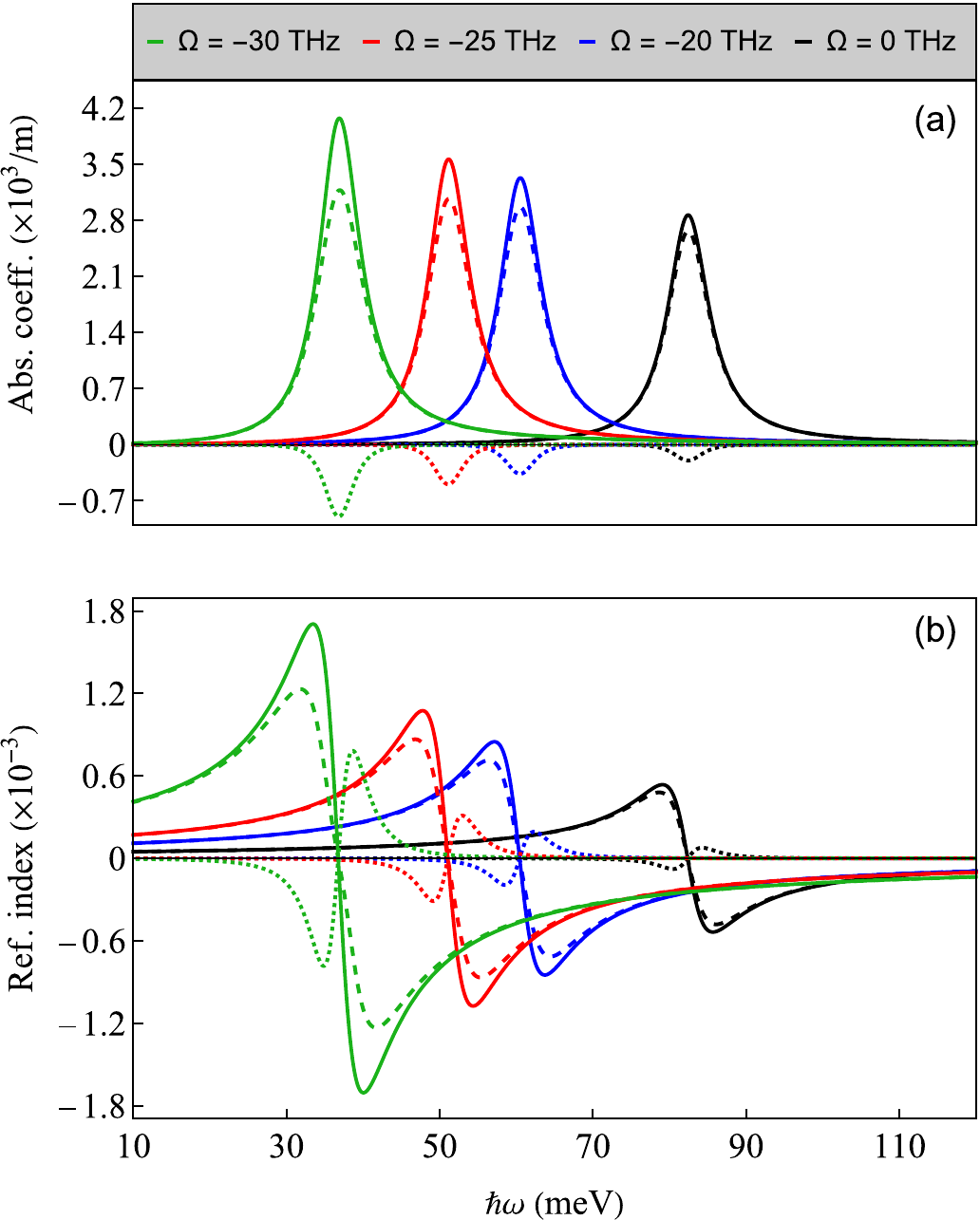}
\caption{The same as in Fig. \ref{f2}, but considering negative rotations. In both Figures, we consider $\phi=0.5\,(h/e)$ and $B=15$ Tesla.}
\label{f2N}	
\end{figure}
being the linear refractive coefficient, and
\begin{equation}
\alpha ^{(3)}\left( \omega ,I\right) =-\hslash \omega \sqrt{\frac{\mu_0}{
\epsilon_{r}}}\frac{4 I}{2\epsilon _{0}n_{r}c}\frac{\sigma _{\nu }\Gamma
_{21}\left\vert M_{21}\right\vert ^{4}}{\left[\left(\Delta E-\hslash
\omega \right) ^{2}+\left( \hslash \Gamma _{21}\right) ^{2}\right]^2}
\end{equation}
is the third-order nonlinear refractive coefficient, where $\epsilon_r=\epsilon _{0}n_{r}^2$ and $I=2\epsilon _{0}n_{r}c\left\vert \mathcal{E}\right\vert ^{2}
$ represents the incident optical intensity.
Similarly, the total refractive coefficient is given by
\begin{equation} 
\frac{\Delta n \left( \omega ,I\right)}{n_r} =\frac{\Delta n^{(1)}\left( \omega\right) }{n_r}+\frac{\Delta n^{(3)}\left( \omega ,I\right)}{n_r} ,  \label{nt}
\end{equation}
where
\begin{equation}
\frac{\Delta n^{(1)}\left( \omega\right) }{n_r} =\frac{\sigma _{\nu }\left\vert M_{21}\right\vert ^{2}}{2 n_r^{2}\epsilon_0}\frac{\Delta E-\hslash\omega}{\left(\Delta E-\hslash \omega \right) ^{2}+\left( \hslash \Gamma
_{21}\right) ^{2}}  \label{n1}
\end{equation}
is the linear refractive coefficient, and
\begin{equation}
\frac{\Delta n^{(3)}\left( \omega ,I\right)}{n_r} =-\frac{ \mu c I \sigma _{\nu }\left\vert M_{21}\right\vert ^{4}}{\epsilon _{0}n_{r}^3}\frac{\Delta E-\hslash\omega}{\left[\left(\Delta E-\hslash
\omega \right) ^{2}+\left( \hslash \Gamma _{21}\right) ^{2}\right]^2} \label{n3}
\end{equation}
is the third-order nonlinear refractive coefficient.

In Figure \ref{f2}, we numerically calculate the OACs and the RICs as a function of photon energy for four different values of the rotation parameter $\Omega$, i.e.,  $\Omega=0$, $\Omega=30 \,\text{THz}$, $\Omega=50\,\text{THz}$ and $\Omega=80 \,\text{THz}$. Our calculations focus on the optics involved in the transition between states with ($n=0,m=0$) and ($n=2, m=1$). Figure \ref{f2}(a) shows the linear, third-order non-linear, and total optical absorption coefficients as a function of incident photon energy for varying rotation values. The linear coefficient dominates, while the third-order coefficient has a small contribution. As the rotation parameter $\Omega$ increases, the resonant peaks (also known as transition frequency) at $\hbar\omega=\Delta E$ shift to higher energies, and the amplitudes of all the OACs decrease.
Figure \ref{f2}(b) shows the linear, third-order non-linear, and total refractive index for varying rotation values. The RICs exhibit the same behavior concerning rotation effects as the OACs, with peak positions shifting to higher energies and resulting in a blue resonance shift in a two-dimensional quantum ring. The peak intensities of all RICs decrease as $\Omega$ increases.
Notably, the RICs are located at the resonant peak $\hbar\omega=\Delta E$ of the OACs, as expected. Furthermore, the influence of the rotating parameter is evident in Figure \ref{f2}(a)-(b), where the curves with rotation effects (colored) are separated from those without rotation effects (black).  

In Figure \ref{f2N}, we also present numerical calculations of the OACs and the RICs as a function of photon energy for the transition between states with ($n=0, m=0$) and ($n=2, m=1$). Here, the calculations are performed for four different values of the rotation parameters, including negative rotations: $\Omega=0$, $\Omega=-20$ THz, $\Omega=-25$ THz, and $\Omega=-30$ THz. The Figure \ref{f2N} reveals a similar trend to what was observed in Figure \ref{f2}. However, it is evident that as the rotation parameters decrease (negative rotations), the effects of non-linear optics become more pronounced (see the third-order non-linear coefficients (dotted
line) in Figure \ref{f2N}).

\section{Conclusions}
\label{sec:conclusions}

In this work, we have investigated the nonrelativistic quantum motion of a charged particle in a 2D quantum ring under the influence of rotation. By deriving the equation of motion using the minimal coupling procedure and considering the Aharonov-Bohm effect and a uniform magnetic field, we obtained the eigenvalues and eigenfunctions, providing insights into the energy levels and spatial distribution of the particle. Our analysis of the probability distribution revealed a distinct shift towards lower values along the radial coordinate as the rotation increased. This shift indicates a higher likelihood of finding the particle closer to the inner and outer radii (edges) of the system rather than in the average region $\rho_0$. These results are consistent with the observed occupation of electrons at the system's edges, emphasizing the singular dependence of electronic states on the rotation parameter. Furthermore, we examined the optical properties of the system, including optical absorption and refractive coefficients. Our findings demonstrated that rotational effects have a significant impact on the system's behavior. In both cases, we observed a shift towards higher values of the resonant energy peak, accompanied by a decrease in the amplitude of the peaks with increasing rotational effects. The insights gained from this study deepen our understanding of the intricate relationship between the rotation and quantum behavior in quantum rings. The observed concentration of electrons at the system's edges and the significant rotational effects on optical properties may contribute to the development of quantum optics, optoelectronics, and related technologies. The investigations presented in this work provide valuable insights into the behavior of charged particles in rotating quantum ring systems, shedding light on the fundamental aspects of quantum mechanics and opening avenues for further exploration.

To conclude, we would like to emphasize that our analysis considers significant rotational effects, typically on the scale of several terahertz (THz). These high rotation rates are necessary to achieve compelling visualizations of the rotational effects, as discussed in \cite{pereiramodification}, when compared to scenarios where rotations are absent. However, it is worth noting that in quantum systems such as 2D quantum rings, rotational effects exist even at much lower rotation rates.

\section*{Acknowledgments}

This work was partially supported
by the Brazilian agencies CAPES, CNPq, and FAPEMA. E. O. Silva acknowledges CNPq
Grant 306308/2022-3, FAPEMA Grants PRONEM-01852/14 and UNIVERSAL-06395/22. F. S. Azevedo acknowledges CNPq Grant No. 150289/2022-7. C. Filgueiras acknowledges FAPEMIG Grant No. APQ 02226/22.
This study was financed in part by the Coordena\c{c}\~{a}o de
Aperfei\c{c}oamento de Pessoal de N\'{\i}vel Superior - Brasil (CAPES) -
Finance Code 001.

\bibliographystyle{apsrev4-2}
%

\end{document}